\documentclass[prc,superscriptaddress,unsortedaddress,twocolumn]{revtex4}
\usepackage[english]{babel}
\usepackage{graphicx,epsf,bm,amsmath,color}
\usepackage{here,ulem}

\def\bp{\mbox{\boldmath $p$}}
\def\bP{\mbox{\boldmath $P$}}
\def\bq{\mbox{\boldmath $q$}}
\def\br{\mbox{\boldmath $r$}}
\def\bR{\mbox{\boldmath $R$}}

\newif\iffigure
\figuretrue
\begin{document}
\title{$\Lambda$-deuteron momentum correlation functions incorporating
deuteron breakup contributions in Faddeev formulation}
\author{M. Kohno}
\affiliation{Research Center for Nuclear Physics, Osaka University, Ibaraki 567-0047,
Japan}

\author{H. Kamada}
\affiliation{Research Center for Nuclear Physics, Osaka University, Ibaraki 567-0047,
Japan}

\begin{abstract}
The effects of the deuteron breakup are estimated for the $\Lambda$-deuteron momentum
correlation function. Faddeev amplitudes in calculating low-energy $\Lambda$-deuteron
scattering can provide not only the elastic scattering part but also breakup wave functions
in the incident and the rearrangement channels. Calculations are carried out using
nucleon-nucleon (NN) and hyperon-nucleon (YN) interactions parametrized in chiral effective
field theory. The effects of the breakup in the incident channel are found to be marginally
insignificant. Those of the rearrangement channel are not negligible, but not large
when the source radius is larger than $2.5$ fm. Nevertheless, it is worthwhile to have
the information on the magnitude of these effects when analyzing the experimental data.
\end{abstract}

\maketitle
\section{Introduction}
A proper description of the baryon-baryon interactions in the strangeness sectors of $S=-1$ and $-2$
is essential for a microscopic understanding of hypernuclei and neutron star matter. Although
progress is being made in direct hyperon-nucleon scattering experiments, e.g., \cite{Miwa21},
data from scattering experiments are currently scarce. Recently another source of information
has become available from the momentum correlation functions between the hyperon and the nucleon
or between the hyperon and the light nuclei that are measured in heavy-ion collision experiments.
These data can be used to constrain the parametrization of the hyperon-nucleon
potentials \cite{MHM24}.
   
We reported in Ref. \cite{KK24} the results of Faddeev calculations of the phase shifts of
the $\Lambda$-deuteron elastic scattering at low energies, using hyperon-nucleon
(NLO13 \cite{NLO13} and NLO19 \cite{NLO19}) and chiral nucleon-nucleon (NN) interactions
(N$^4$LO$^+$ \cite{RKE18}) interactions. The $\Lambda$-deuteron relative wave function
calculated from the $\Lambda d$ $T$-matrices has been used to evaluate the relevant correlation
functions. The results obtained by the relative wave function of the Faddeev calculations
agree well with that of the Lednicky-Lyuboshits (LL) formula \cite{LL82} using the effective
range parameters obtained by the Faddeev calculations, when the source radius is larger
than $\sim 1$ fm. This indicates that the LL formula is useful to analyze experimental data.

Because the deuteron is not an elementary particle, a process in which the deuteron is produced
from the general $\Lambda np$ state is possible. The consideration of such a process corresponds
to including the deuteron breakup wave functions.  Although a modification of the deuteron wave
function in the interaction region is taken into account in the Faddeev calculations,
there are other breakup processes in both the incident and rearrangement channels.
These breakup components can be provided by the Faddeev amplitudes. It is desirable
to study these contributions and to estimate the order of the magnitude of their effects
for the analysis of the experimental data of the $\Lambda d$ momentum correlation function,
to obtain information about the $\Lambda N$ interaction.

In this article, we take into account the deuteron breakup processes in evaluating the $\Lambda d$
correlation function to study these effects. The correlation functions involving composite particles
have been studied by Mr\'{o}wczy\'{n}ski \cite{Mr20}. We follow his derivation of the theoretical
correlation function. Its expression is recapitulated for the sake of completeness in Sec. II.  
The evaluation of the full $\Lambda d$ correlation function requires $\Lambda np$ breakup
wave functions in addition to the two-body $\Lambda d$ relative wave function. Section III
summarizes the simultaneous Faddeev equations for the $\Lambda d$
scattering used in Ref. \cite{KK24}, and then  presents the method of obtaining the three-body
wave functions in configuration space from the Faddeev amplitudes in momentum space
by a spectral representation of the Green function. The three-body wave function in the
incident channel is examined separately from the full three-body wave function,
which includes that of the rearrangement channel. Calculated results of the $\Lambda d$
correlation functions are shown in Sec. IV. A summary follows in Sec. V.

\section{correlation function including three-body dynamics}
The deuteron ($d$) is a bound state of a neutron ($n$) and a proton ($p$). To obtain the
theoretical $\Lambda d$ momentum correlation function, the formulation by
Mr\'{o}wczy\'{n}ski \cite{Mr20} is followed. For the sake of brevity, the following expressions
do not include an indication of the spin and isospin degrees of freedom.
The momentum correlation function, denoted by $C$, is defined as
\begin{align}
 C(q_0)A_2=& g_sg_I (2\pi)^3 \int d \br_\Lambda d \br_n d \br_p\; D(r_n)D(r_p)D(r_\Lambda)
 \notag \\
 &\times |\psi_{\Lambda np}(\br_\Lambda,\br_n,\br_p;q_0)|^2,
\label{eq:def}
\end{align}
where $\psi_{\Lambda np}(\br_\Lambda,\br_n,\br_p;q_0)$ is a three-body wave function with
$q_0$ being the asymptotic relative momentum between $\Lambda$ and $d$, $g_s g_I=3/4$
is a statistical spin-isospin factor for the deuteron formation, and $D(r)$ is a Gaussian source
function with the range parameter $R_s$:
\begin{equation}
 D(r)=D(r;R_s)\equiv (2\pi R_s^2)^{-3/2}\exp\{-r^2/(2R_s^2)\}.
\end{equation}
Although the source radius can be different between the nucleon and the $\Lambda$ hyperon,
the same radius $R_s$ is assumed in the present article.
$A_2$ is a deuteron formation rate given by
\begin{equation}
 A_2= g_sg_I (2\pi)^3 \int d \br D_{np}(r) |\phi_d(\br)|^2,
\label{eq:a2}
\end{equation}
where $\phi_d(\br)$ is a deuteron wave function and the source function $D_{np}(r)$ for the
deuteron is obtained by the convolution of the source function $D(r)$ as
\begin{align}
 & D_{np}(r)=\int d \bR \;D(|\bR+\frac{1}{2}\br|)D(|\bR-\frac{1}{2}\br|) \notag \\
 & =(4\pi R_s^2)^{-3/2}\exp\{-r^2/(4R_s^2)\}=D(r;\sqrt{2}R_s).
\label{eq:sfn}
\end{align}
To remove the center-of-mass degrees of freedom in Eq. (\ref{eq:def}),
the Jacobi spatial coordinates $(\bR,\br_{np},\br_3)$ are introduced as follows,
assuming the same mass for $n$ and $p$. Denoting the nucleon and $\Lambda$ masses
as $m_N$ and $m_\Lambda$, respectively, these coordinates are expressed as
\begin{align}
 & \bR \equiv \frac{m_N \br_n +m_N \br_p +m_\Lambda\br_\Lambda}{M},\; \br_{np}
 \equiv \br_n-\br_p,\notag \\
 & \br_{3} \equiv \br_\Lambda-\frac{1}{2}(\br_n +\br_p),
\end{align}
where $M\equiv 2 m_N+m_\Lambda$. The original coordinates are given as
\begin{align}
 & \br_n =\bR-\frac{m_\Lambda}{M}\br_{3}+\frac{1}{2}\br_{np},\;
  \br_p =\bR-\frac{m_\Lambda}{M}\br_{3}-\frac{1}{2}\br_{np},\notag \\
 &  \br_\Lambda=\bR+\frac{2 m_N}{M} \br_{3},
\end{align}
and the Jacobian of the coordinate transformation is 1:
\begin{equation}
 d\br_\Lambda d\br_n d\br_p= d\bR d \br_{np} d \br_{3},
\end{equation}
The free center-of-mass motion of the three-body system, the momentum of which is
denoted by $\bP$, is separated and disappears when taking the absolute value in Eq. (\ref{eq:def}):
\begin{equation}
 \psi_{\Lambda np}(\br_\Lambda,\br_n,\br_p)=e^{i\bP\cdot\bR} \psi(\br_{np},\br_{3}).
\end{equation}
Then, noticing
\begin{equation}
 r_n^2+r_p^2+r_\Lambda^2= 3\left(\bR -\frac{2(m_\Lambda-m_N)}{3M}\br_{3}\right)^2
 +\frac{2}{3}\br_{3}^2 +\frac{1}{2}\br_{np}^2,
\end{equation}
the $d\bR$ integration is analytically carried out in Eq. (\ref{eq:def}) and
the factor $R(r_n)R(r_p)R(r_\Lambda)$ in Eq. (\ref{eq:def}) is reduced to
\begin{align}
 & \int d\bR\;R(r_n)R(r_p)R(r_\Lambda) \notag \\
=& (3\pi R_s^2)^{-3/2} \exp \left\{-\frac{\br_{3}^2}{3R_s^2}
 \right\}(4\pi R_s^2)^{-3/2}\exp\left\{-\frac{r_{np}^2}{4R_s^2}\right\}\notag \\
 =& D(r_3;\sqrt{3/2}R_s) D(r_{np};\sqrt{2}R_s) \equiv  D_{\Lambda d}(r_{3})D_{np}(r_{np}) .
\label{eq:sf}
\end{align}
Using Eq. (\ref{eq:a2}), Eq. (\ref{eq:def}) becomes
\begin{align}
 C(q_0)=\frac{\int d \br_{3} d \br_{np}\;D_{\Lambda d}(r_{3})D_{np}(r_{np})|\psi(\br_{np},\br_{3})|^2}
 {\int d \br D_{np}(r_{np})|\phi_d(\br)|^2}.
\label{eq:defr}
\end{align}

The wave function $\psi(\br_{3},\br_{np})$ can be expanded in partial waves in each
coordinate:
\begin{align}
 \psi(\br_{np},\br_{3})&=\sum_{\ell,\lambda}(2\ell+1)(2\lambda+1)i^{\ell+\lambda}
 P_\ell(\cos\widehat{\bp_0 \br_{np}}) \notag \\
 & \times P_\lambda(\cos\widehat{\bq_0 \br_{3}}) \phi_\ell(r_{np};p_{q_0})
 \varphi_\lambda(r_{3};q_0 ),
\end{align}
where $P_\ell$ stands for a Legendre polynomial and $\bq_0$ ($\bp_0$) is an asymptotic
momentum in the spatial coordinate $\br_3$ ($\br_{np}$).
If the $\Lambda$ hyperon does not interact with the deuteron, the wave function between
$\Lambda$ and $d$ is a plane wave. In this case, the wave function is represented as
\begin{align}
 & \psi_0 (\br_{np},\br_{3}) = \phi_d (\br_{np}) e^{i\bq_0 \cdot\br_3} \notag \\
 &= \phi_d (\br_{np}) \sum_{\lambda}(2\lambda+1)i^{\lambda}
 P_\lambda(\cos \widehat{\bq_0\br_{3}})  j_\lambda(q_0 r_{3}),
\end{align}
where $j_\lambda$ is a spherical Bessel function.
\begin{widetext}
Noticing that $|\psi_0 (\br_{3},\br_{np})|=|\phi_d (\br_{np})|$,
it is convenient to rewrite Eq. (\ref{eq:defr}) as
\begin{align}
 C(q_0)=  1+\frac{\int d \br_{3} d \br_{np}\; D_{\Lambda d}(r_{3})D_{np}(r_{np})
 \{|\psi(\br_{np},\br_{3})|^2-|\psi_0(\br_{np},\br_{3})|^2\}}{\int d \br D_{np}(r_{np})|\phi_d(\br)|^2}.
\label{eq:cor1}
\end{align}
It is reasonable in low-energy scattering to expect that partial waves other than the $s$ wave
are hardly affected by the interaction:
\begin{equation}
|\psi(\br_{np},\br_{3})|^2-|\psi_0(\br_{np},\br_{3})|^2 \rightarrow |\sum_{\ell=0,2} (2\ell+1)i^\ell
 P_\ell(\cos\widehat{\bp \br_{np}}) \phi_\ell(p_{q_0}r_{np})
 \varphi_0(r_{3};q_0)|^2-|\phi_d(\br_{np})|^2 | j_0(q_0r_{3})|^2.
\label{eq:cor2}
\end{equation}
\end{widetext}
In the case where the deuteron is supposed to be an elementary particle, the following
replacement is applicable:
\begin{equation}
 \sum_{\ell=0,2} (2\ell+1)i^\ell P_\ell(\cos\widehat{\bp \br_{np}})  \phi_\ell(p_{q_0}r_{np})
 \rightarrow \phi_d(\br_{np}). \label{eq:d}
\end{equation}
Then, the correlation function reduces to the frequently employed form of
\begin{equation}
 C(q_0)=1+\int d \br\;  D_{np}(r_{np}) (|\varphi_0(r;q_0)|^2-| j_0(q_0r)|^2).
\end{equation}
The further assumption that $\varphi_0(r;q_0)$ is approximated by its asymptotic form
described by the effective range parameters, the scattering length $a_s$ and the effective
range $r_e$, leads to the LL formula \cite{LL82}:
 \begin{align}
 C(q_0) \approx &\; 1+\frac{|f_J(k)|^2}{2R_s^2} F(r_e)
  +\frac{2\mbox{Re}f_J(k)}{\sqrt{\pi}R_s}F_1(x) \notag \\
 & -\frac{\mbox{Im}f_J(k)}{R_s}F_2(x),
\label{eq:app}
\end{align}
where $x\equiv 2kR_s$ and $f_J$ is a scattering amplitude that is approximated
by the effective range parameters as
\begin{equation}
 f_J= \frac{1}{k\cot\delta_J -ik}
 \approx \frac{1}{-\frac{1}{a_s}+\frac{1}{2}r_e k^2-ik}.
\end{equation}
$R_s$ is the source range.  Three functions $F$, $F_1$,
and $F_2$ are given by $F(r_e)=1-r_e/(2\sqrt{\pi}R_s)$,
$F_1(x)= \int_0^x dt\:e^{t^2-x^2}/x$, and $F_2(x)=(1-e^{-x^2})/x$, respectively.

In this article, we do not adopt the assumption of Eq. (\ref{eq:d}), but employ the expression of
Eqs. (\ref{eq:cor1}) and (\ref{eq:cor2}) for the $\Lambda d$ momentum correlation function
using three-body wave functions obtained by Faddeev calculations, in which the wave functions
corresponding to the deuteron breakup process are incorporated in addition to the elastic ones.

\section{three-body wave function}
The evaluation of the momentum correlation functions given by Eqs. (\ref{eq:cor1}) and
(\ref{eq:cor2}) requires a three-particle wave function that is composed of the incident wave
function, the $\Lambda d$ elastic state, and the deuteron breakup state. The $\Lambda d$
elastic scattering wave function was calculated in Ref. \cite{KK24} to evaluate the $\Lambda d$
momentum correlation function. The breakup wave functions manifest in two distinct channels:
the incident channel and the rearrangement channel. These wave functions are derived from
the solution of the Faddeev equations, as explained below. As outlined in Ref. \cite{KK24},
the Faddeev equations for the $\Lambda d$ scattering are formulated as simultaneous equations
for the two operators $T_2$ and $T_3$:
\begin{align}
 \langle p_3 q_3 \alpha_3|T_3|\phi\rangle =& \langle p_3 q_3 \alpha_3|t_3 G_0 (1-P_{12})T_2
 |\phi\rangle, \label{eq:pe1}\\
 \langle p_2 q_2 \alpha_2|T_2|\phi\rangle =& \langle p_2 q_2 \alpha_2|t_2 + t_2 G_0 T_3
 - t_2 P_{12} G_0  T_2 |\phi\rangle,
\label{eq:pe2}
\end{align}
where $|\phi\rangle=|\phi_d, q_0\rangle$ is an incident wave function, $G_0$ is a three-body
free Green function, and $t_i$ is a pertinent two-body $t$-matrix.
Two sets of the Jacobi momenta $(\bp_j,\bq_j)$ with $j=2,3$ are defined as in Fig. 1.
The indicator $\alpha_i$ is used to specify the partial-wave channel. The $\Sigma$ hyperon
is not  included in the basis states, although the $\Lambda N$-$\Sigma N$ coupling is
taken care of in evaluating the two-body $t_2$-matrix.

The spatial coordinates corresponding to the Jacobi momenta $(\bp_3,\bq_3)$ are denoted by
$(\br_{np},\br_3)$. The three-body wave functions in the configuration space are derived
from the obtained matrix elements of the operators $T_3$ and $T_2$ in momentum space.

\begin{figure}[b]
\centering
\includegraphics[width=0.3\textwidth]{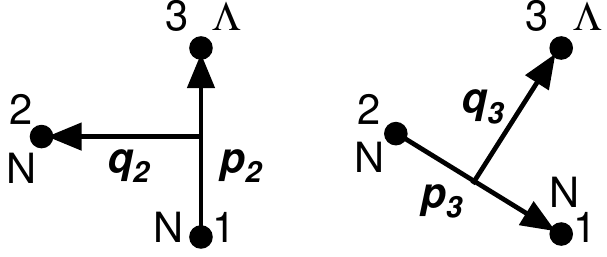}
\caption{Two sets of Jacobi momenta.}
\label{fig:jacobi}
\end{figure}

\subsection{Incident channel wave function}
The incident channel wave function in the configuration space is represented by \cite{GL96}
\begin{align}
 &  \psi(\br_{np},\br_{3})= \langle \br_{np},\br_3|\Psi_3^{(+)}\rangle \notag \\
 & = \langle \br_{np},\br_3|\phi\rangle
  +\langle \br_{np},\br_3|G_3(V_{23}+V_{31})|\Psi_3^{(+)}\rangle \notag \\
 &= \langle \br_{np},\br_3|\phi\rangle +\langle \br_{np},\br_3|G_3 (1-P_{12})T_2|\phi\rangle,
\label{eq:chwf}
\end{align}
where $G_3$ is the Green function of the incident channel Hamiltonian $H_0+V_{12}$
for the total energy $E$:
\begin{equation}
 G_3= [E-H_0-V_{12}+i\epsilon]^{-1}.
\end{equation}
Since there is a deuteron bound state for the interaction $V_{12}=V_{np}$, the spectral
representation of $G_3$ in the pertinent channel is expressed as follows:
\begin{align}
 G_3=&\int d\bq \frac{|\phi_d, \bq\rangle \langle \phi_d,\bq|}
{E-e_d-\frac{\hbar^2}{2\mu_{\Lambda NN}}\bq^2 +i\epsilon} \notag \\
 &+\iint d\bp d\bq \frac{|\psi_{\bp}, \bq\rangle \langle \psi_{\bp},\bq|}
{E-\frac{\hbar^2}{2\mu_{NN}}\bp^2-\frac{\hbar^2}{2\mu_{\Lambda NN}}\bq^2 +i\epsilon},
\label{eq:g3}
\end{align}
where $\phi_d$ is the wave function of the deuteron bound state with an energy of $e_d$,
$\mu_{\Lambda NN}$ is a reduced mass of
$\mu_{\Lambda NN}=\frac{2m_N m_\Lambda}{2m_N+m_\Lambda}$,
$|\bq\rangle$ is a plane wave with the momentum $\bq$, and $|\psi_{\bp}\rangle$
is a scattering wave function that satisfies the Schr{\"o}dinger equation
\begin{align}
 \left(-\frac{\hbar^2}{2\mu_{NN}} \hat{\bp}_3^2 +V_{np} \right)\psi_{\bp}(\bp_3)
  =\frac{\hbar^2 \bp^2}{2\mu_{NN}}\psi_{\bp}(\bp_3),
\end{align}
with $\mu_{NN}=m_N/2$.
In the calculations of the correlation function, a configuration space representation is required for
$|\psi_{\bp}\rangle$. On the other hand, an expression in momentum space is needed for
$\langle \psi_{\bp}|$, because the Faddeev amplitude $T_2$ is solved in momentum space.
The wave function in momentum space is given as a generalized function comprising
a $\delta$ function and the kernel of a principal value integration.
The explicit forms of these wave functions in both configuration space and momentum space
in the case of an $s$ wave are provided in the Appendix. 

Substituting Eq. (\ref{eq:g3}) into Eq. (\ref{eq:chwf}), the first term of Eq. (\ref{eq:g3})
yields the elastic scattering wave function. This contribution was evaluated in Ref. \cite{KK24}.
The second term of Eq. (\ref{eq:g3}) is the wave function corresponding to the deuteron breakup
other than the modification of the deuteron intrinsic wave function in the initial channel.
It is important to note that $|\psi(\bp)\rangle$ is normalized by the factor of $(2\pi)^{-3/2}$
corresponding to the bound deuteron wave function. On the other hand, $|\bq\rangle$ does not
accompany this normalization due to the definition of the correlation function.   

In the subsequent calculations of the correlation functions, the $\ell=0$ part is
considered exclusively, and the $s$-state deuteron wave function is employed in Eq. (\ref{eq:cor2}),
although the $\ell=2$ states are included in both $NN$ and $\Lambda N$ channels in
solving the Faddeev equations. The exclusion of the $d$ state does not alter the
estimation of the order of the magnitude of the deuteron breakup effect for
the $\Lambda$-deuteron correlation function.
Given that the matrix elements of the $T_2$ operator
are obtained in the $(p_2,q_2,\alpha_2)$ Jacobi basis, the transformation of
the Jacobi basis to $(p_3,q_3,\alpha_3)$ is necessary, employing the recoupling coefficients
$\langle p_3,q_3,\alpha_3|p_2,q_2,\alpha_2\rangle$. The explicit expression of
Eq. (22) is as follows:
\vspace*{-6mm}
\begin{widetext}
\begin{align}
 & \psi_{\ell=0}(r_{np},r_3) \notag  \\=& \phi_d(r_{np}) \left\{ j_0(q_0 r_3)
 + 2\iint p_3^2 dp_3 \:q_3^2 dq_3 \frac{\tilde{\phi}_d (p_3) j_0(q_3 r_3)}
{E-e_d-\frac{\hbar^2}{2\mu_{\Lambda d}}q_3^2 +i\epsilon} \iint p_2^2 dp_2 \:q_2^2 dq_2
 \langle p_3,q_3,\alpha_3|p_2,q_2,\alpha_2\rangle \langle p_2,q_2,\alpha_2|T_2|\phi\rangle \right\}
 \notag \\
  +&  2\int p^2 dp \iint p_3^2 dp_3\: q_3^2 dq_3  \iint p_2^2 dp_2 \:q_2^2 dq_2
 \frac{\psi_{\ell=0}(r_{np};p) \tilde{\psi}_{np}(p_3;p) j_0(q_3 r_3)}
{E-\frac{\hbar^2}{2\mu_{NN}}p^2-\frac{\hbar^2}{2\mu_{\Lambda NN}}q_3^2 +i\epsilon}
  \langle p_3,q_3,\alpha_3|p_2,q_2,\alpha_2\rangle \langle p_2,q_2,\alpha_2|T_2|\phi\rangle.
\end{align}
\end{widetext}

\subsection{Wave function including rearrangement breakup}
The full three-body wave function is given by the Faddeev amplitudes as follows \cite{GL96}:
\begin{align}
 & \psi(\br_{np},\br_{3})= \langle \br_{np},\br_3|\Psi_3^{(+)}\rangle \notag \\
 & =\langle \br_{np},\br_3|G_0 (V_{12}+V_{23}+V_{31})
 |\Psi_3^{(+)}\rangle \notag \\
 &= \langle \br_{np},\br_3|G_0 \{G_0^{-1}+(1-P_{12})T_2+T_3\}|\phi\rangle.
\label{eq:fwf}
\end{align}
$G_0$ is a free three-body Green function:
\begin{align}
 G_0=&[E-H_0+i\epsilon]^{-1} \notag \\
 =& \iint d\bp_3 d\bq_3 \frac{|\bp_3, \bq_3\rangle \langle \bp_3,\bq_3|}
{E-\frac{\hbar^2}{2\mu_{NN}}\bp_3^2-\frac{\hbar^2}{2\mu_{\Lambda NN}}\bq_3^2 +i\epsilon},
\end{align}
where $|\bp_3,\bq_3\rangle$ is a plane wave both in $\bp_3$ and $\bq_3$ momenta.
This wave function contains the deuteron breakup effects in both the initial and rearrangement
channels. It is important to reiterate that the natural normalization factor $(2\pi)^{-3/2}$ is
assigned for $|\bp_3\rangle$, yet it is not assigned for $|\bq_3\rangle$.
 
To proceed with the numerical calculation, it is necessary to obtain the following matrix element:
\begin{equation}
 \langle p_3q_3\alpha_3|2T_2+T_3 |\phi\rangle,
\end{equation}
This matrix element requires specific treatments due to the presence of a bound-state pole in the
case where $\alpha_3$ is the deuteron channel. That is, the matrix element in the relevant channel
is processed as
\begin{equation}
 \langle p_3q_3\alpha_3|T_3|\phi\rangle=\frac{\langle p_3q_3\alpha_3|\tilde{T}_3|\phi\rangle}
{q_0^2-q^2+i\epsilon},
\end{equation}
where $q_0=\{2\mu_{\Lambda NN}E_{cm}/\hbar^2\}^{1/2}$ for the center-of-mass energy
$E_{cm}=E-e_d$.
The explicit form of Eq. (\ref{eq:fwf}) for the $s$ wave is as follows. The matrix elements of the
$T_3$ operator are obtained in the $(p_3,q_3,\alpha_3)$ Jacobi basis, the transformation of
the Jacobi basis is not necessary for the $T_2$ term:

\begin{widetext}
\begin{align}
 & \psi_{\ell=0}(r_{np},r_3) = \phi_d(r_{np}) j_0(q_0 r_3)
 + \iint p_3^2 dp_3 \: q_3^2 dq_3 \frac{j_0(p_3 r_{np}) j_0(q_3 r_3)}
{E-\frac{\hbar^2}{2\mu_{NN}}p_3^2-\frac{\hbar^2}{2\mu_{\Lambda NN}}q_3^2 +i\epsilon}
 \langle p_3,q_3,\alpha_3|T_3|\phi\rangle  \notag \\
 & \hspace{2em}+ 2\iint p_3^2 dp_3 \:q_3^2 dq_3  \iint p_2^2 dp_2\: q_2^2 dq_2
 \frac{j_0(p_3 r_{np}) j_0(q_3 r_3)}
{E-\frac{\hbar^2}{2\mu_{NN}}p_3^2-\frac{\hbar^2}{2\mu_{\Lambda NN}}q_3^2 +i\epsilon}
  \langle p_3,q_3,\alpha_3|p_2,q_2,\alpha_2\rangle \langle p_2,q_2,\alpha_2|T_2|\phi\rangle.
\label{eq:sldwf}
\end{align}
\end{widetext}

\section{Results}
\begin{figure}[b]
\centering
\includegraphics[width=0.4\textwidth]{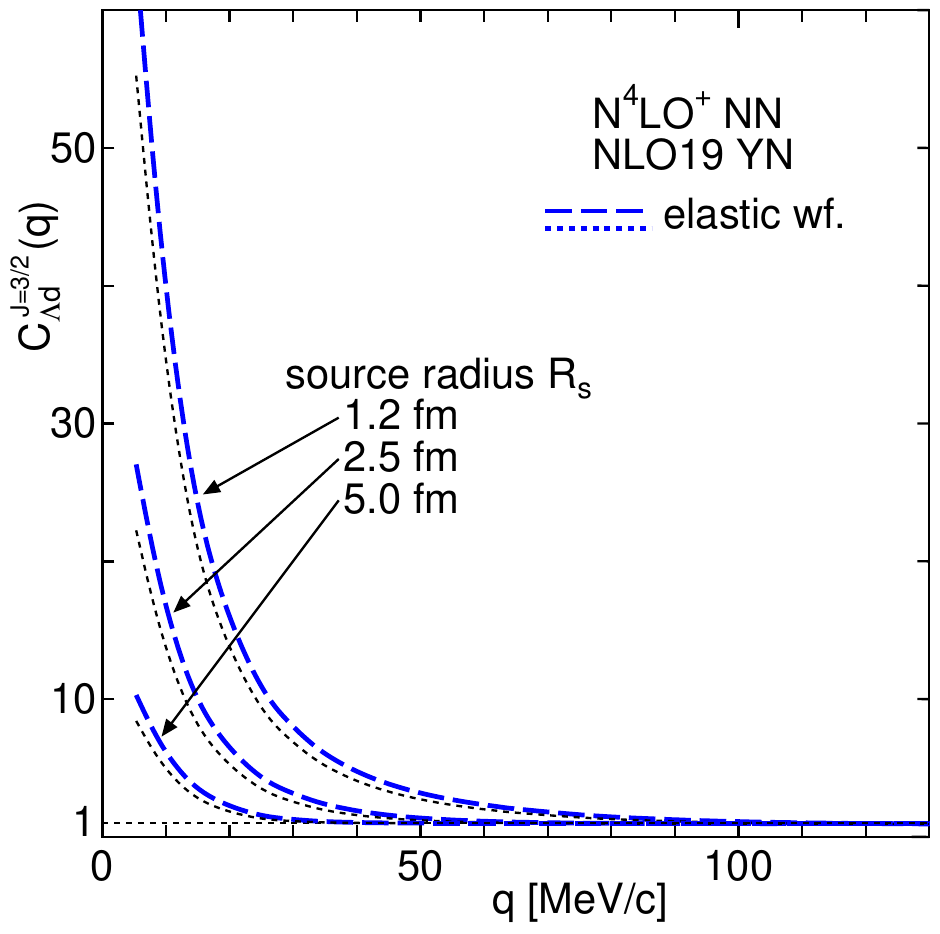}
\caption{Correlation functions evaluated by the $\Lambda d$ relative elastic wave
functions for $J=3/2$ with the N$^4$LO$^+$ NN  \cite{RKE18} and NLO19
YN \cite{NLO19} interactions. The thin dotted curves are the same as those given
in Fig. 6 in Ref. \cite{KK24} in which the source function
$D(r_{np};\sqrt{2}R_s)D(r_{\Lambda d};\sqrt{2}R_s)$ is used.
The thick dashed curves are the results calculated with the source function
given in Eq. (10): $D(r_{np};\sqrt{2}R_s)D(r_{\Lambda d};\sqrt{3/2}R_s)$.}
\label{fig:raddep}
\end{figure}

We use chiral SMS interaction at order N$^4$LO$^+$ \cite{RKE18} for the $NN$ interaction,
and chiral NLO13 \cite{NLO13} and NLO19 \cite{NLO19} realizations of the chiral interactions
at order NLO for the $\Lambda(\Sigma) N$-$\Lambda(\Sigma) N$ interactions.
Prior to the presentation of the contributions of the deuteron breakup wave functions, we show the 
dependence of the momentum correlation function on the choice of the source radius.
As is demonstrated in Eq. (\ref{eq:defr}), the source radius for the
$\Lambda$-deuteron pair differs from that for the $pn$ pair of the deuteron.
In Ref. \cite{KK24}, the $\Lambda d$
correlation functions with the elastic $\Lambda$-deuteron relative wave function were
calculated by assigning a radius of $\sqrt{2}R_s$ instead of $\sqrt{3/2}R_s$ in the notation
of Eq. (\ref{eq:sfn}). The former choice results in the effective assignment of a larger
source radius in comparison with the proper choice. The differences in
the calculated correlation functions with the different radii are shown
in Fig. \ref{fig:raddep} for $R_s=1.2$, 2.5, and 5.0 fm, respectively.
The figure illustrates the effect of the 15\% rescaling of the source radius.

First, we discuss the effect of the breakup in the incident channel.
It is found that the contribution from the second term of the Green function $G_3$
is small in both $J=1/2$ and $J=3/2$ for each source radius $R_s$ adopted.
In the process of $\Lambda d$ scattering, the total isospin is $T=0$.
Consequently, the transition to the isospin singlet $^1$S$_0$ $np$ pair
does not contribute to the $\Lambda np$ system starting from the $\Lambda$-deuteron
incident state. It is expected that the scattering
$^3$S$_1$ wave function, which is orthogonal to the deuteron component, has minimal influence
on the total wave function at low energies. The breakup contributions in the incident channel
are not included in the figures presented below due to their small size.

\begin{figure}[t]
\centering
\includegraphics[width=0.4\textwidth]{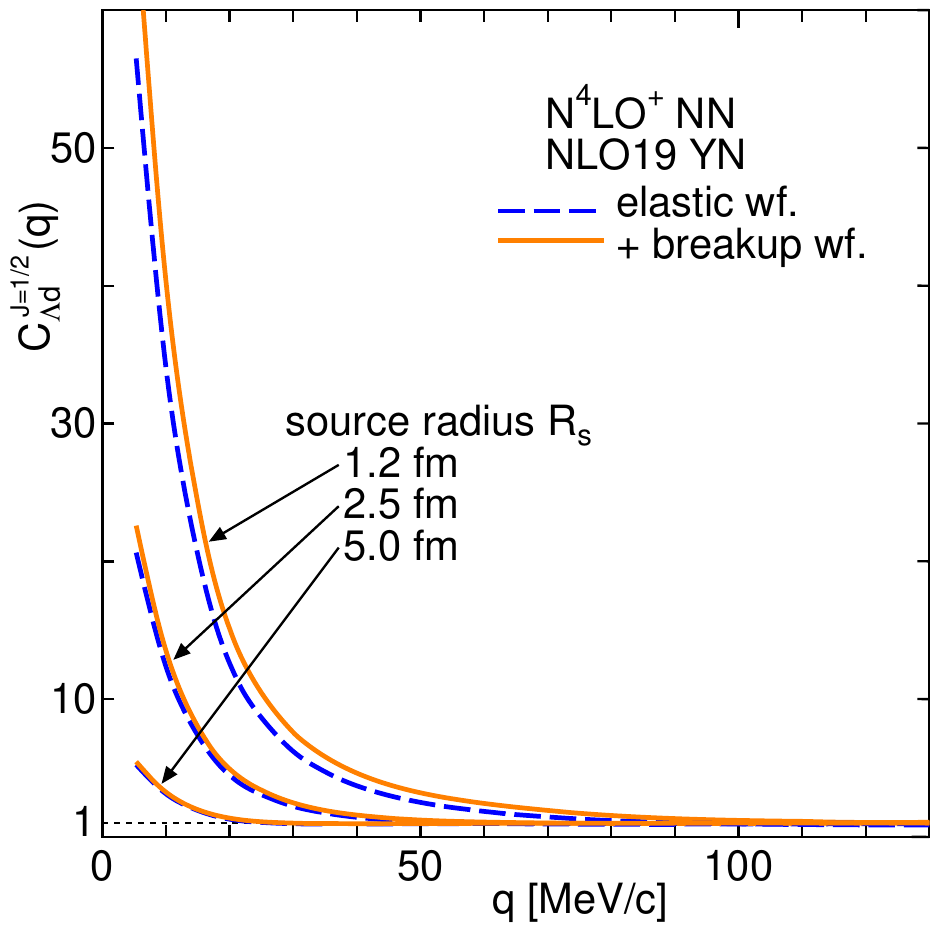}
\caption{The solid curves show the correlation functions evaluated by the full $\Lambda np$
three-body wave function for the $J=1/2$ channel with the N$^4$LO$^+$ NN and NLO19 YN
interactions. The NLO13 YN interaction provides almost same results.
The dashed curves are calculated only with the elastic part. The source function
of Eq. (\ref{eq:sf}) is used with the three cases of $R_s=1.2, 2.5$, and $5.0$ fm.}
\label{fig:19j1}
\end{figure}

Next, the results of the calculation employing the full $\Lambda np$ three-body wave
function of Eq. (\ref{eq:sldwf}) are presented. This wave function incorporates the breakup effect
in the rearrangement channel in addition to that in the initial channel. The correlation functions
are evaluated by Eq. (\ref{eq:cor1}) by assuming that only the $s$ wave deviates from
the plane wave. The results with the NLO19 YN interactions \cite{NLO19} for the $J=1/2$
state are shown in Fig. \ref{fig:19j1}. Because the $J=1/2$ channel is constrained by the
reproduction of the bound hypertriton state, the NLO13 and NLO19 interactions predict
almost same results. In this case, the breakup effect is not negligible.
The effect of the breakup in the rearrangement channel enhances the correlation function.
The magnitude of the observed change is comparable to effect of the rescaling of the source
radius shown in Fig. \ref{fig:raddep}. While the deviation from the elastic correlation function
is small for $R_s > 2.5$ fm, it is important to consider the breakup effect when interpreting
the experimental data.

\begin{figure}[t]
\centering
\includegraphics[width=0.4\textwidth]{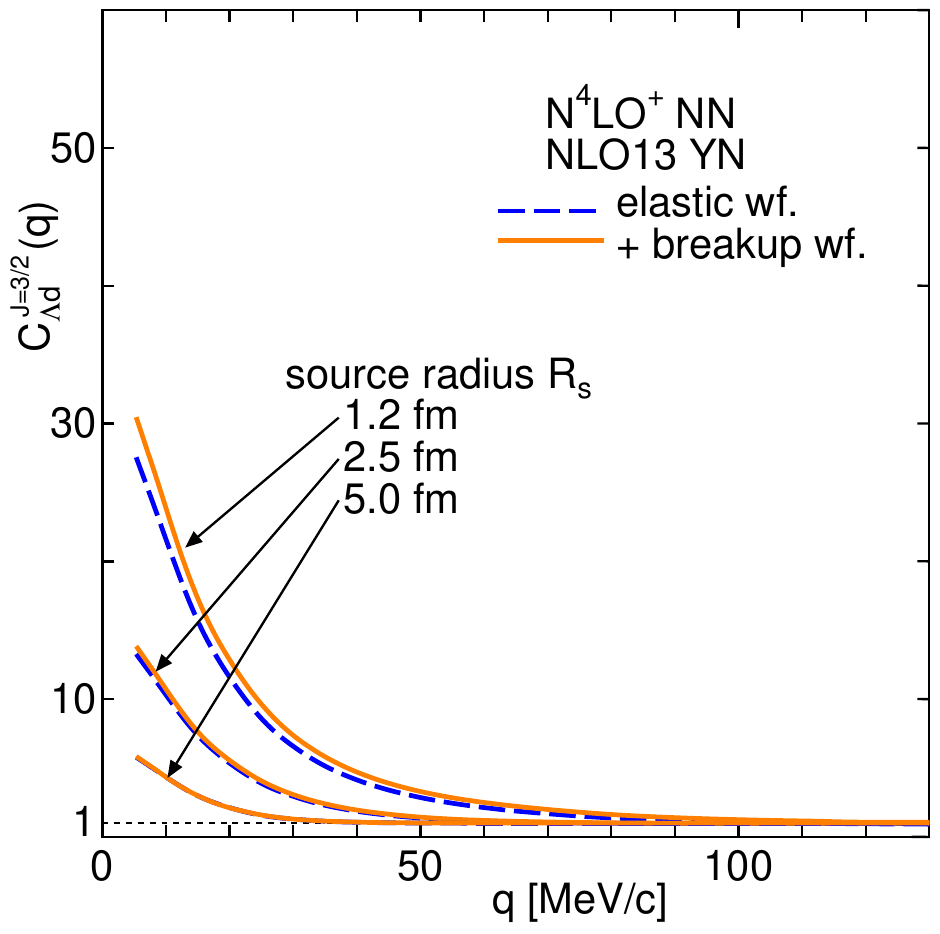}
\includegraphics[width=0.4\textwidth]{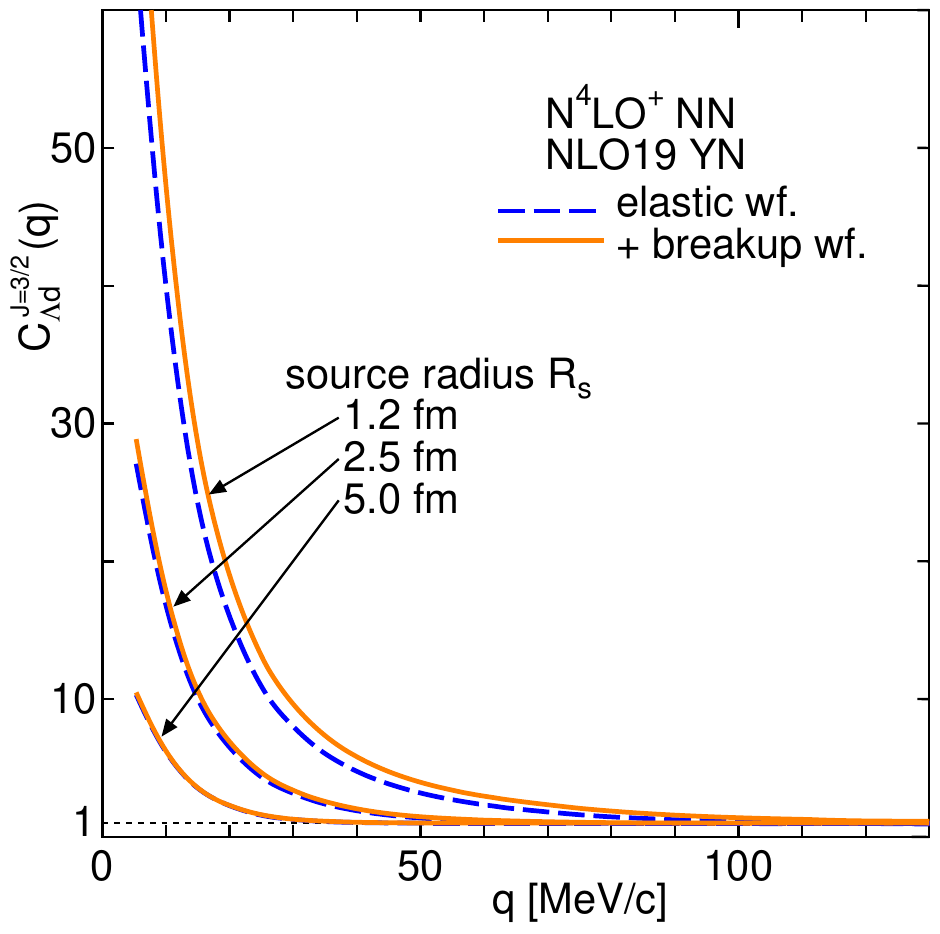}
\caption{Same as Fig. 3, but for the $J=3/2$ channel. The upper (lower) panel shows the results
with the NLO13 (NLO19) YN interactions.}
\label{fig:j3}
\end{figure}

The results for the $J=3/2$ state are shown in Fig. \ref{fig:j3} for NLO13 and NLO19, respectively.
The phase shifts of the $\Lambda d$ elastic scattering presented in Fig. 3 of Ref. \cite{KK24}
showed that NLO19 is more attractive than NLO13 in the $J=3/2$ channel. This feature is
reflected as the difference in the magnitude of the correlation functions.
The magnitude of the correlation function with NLO19 is observed to be approximately
twice as large as that with NLO13, although the difference in the phase shifts shown
in the lower panel of Fig.3 in Ref. \cite{KK24} is seemingly not so large.
This finding indicates that the relative strength between the singlet and triplet
$\Lambda N$ interactions, which is not well determined by the hypertriton binding energy,
can be inferred from the $\Lambda d$ correlation data. 
The contribution of the breakup is small for $R_s > 2.5$ fm as in the case of $J=1/2$.

\section{Summary}
We have considered the effects of deuteron breakup on the $\Lambda d$ momentum correlation
function, as an extension of the previous investigation of $\Lambda$-deuteron low-energy
scattering in the Faddeev formulation \cite{KK24}. In Ref. \cite{KK24}, the $\Lambda d$
correlation functions were evaluated using the $\Lambda d$ elastic wave function, which
was calculated with the half-off-shell $t$-matrices from the Faddeev calculation.
A comparison of these results with the LL formula values, for which the effective range
parameters were those extracted from the Faddeev calculation phase shifts, demonstrated
that the correlation functions obtained in the two descriptions are nearly equivalent
when the source radius is larger than approximately $2.5$ fm.

While the $\Lambda d$ elastic wave functions include the effects of modification of
the deuteron wave function in the scattering process, other effects from the deuteron
breakup must be considered when analyzing the correlation function. The Faddeev
amplitudes can provide contributions from the deuteron breakups in both the incident
channel and in the rearrangement channel,
Consequently, it is worthwhile to estimate these effects. 

The theoretical expression of the $\Lambda$-deuteron correlation function is explained
in Sec. II, following the derivation by Mr\'{o}wczy\'{n}ski \cite{Mr20}.
As indicated in Ref. \cite{Mr20}, the source
radius should differ from that of the two-particle correlation function. Subsequently,
we provided the expression of the wave function including the deuteron breakup processes.
To obtain the wave function in the incident channel  it is necessary to consider the Green function
of the incident channel Hamiltonian in which the interaction between $p$ and $n$ is incorporated.
The explicit expressions of the scattering wave functions in both configuration and momentum
spaces, which are necessary to calculate the Green function, are presented in {the appendix}.
The total three-body wave function, which incorporates the deuteron breakup wave
processes in the rearrangement channel as well as that in the incident channel,
can be obtained using the Green function of the free Hamiltonian.

The effects of the deuteron breakup are discussed after demonstrating how the source function
of $D(r_{\Lambda d};\sqrt{3/2}R_s)$ instead of $D(r_{\Lambda d};\sqrt{2}R_s)$ can enhance
the calculated correlation functions. It is found that the deuteron breakup in
the incident channel gives a negligible effect even for the small source radius. This result is
understandable because the $np$ singlet channel is not involved, and the triplet
breakup wave function has to be orthogonal to the elastic wave function. On the other hand,
the breakup in the rearrangement channel is not negligible.
This result is informative for analyzing experimental data, although the influence of
the selection of the source radius is appreciable. The different magnitudes
of the calculated $\Lambda d$ correlation functions  depending on the NLO13 and NLO19
YN interactions indicate that the future experimental data of the $\Lambda d$ correlation
function is helpful to obtain the relative strength of the $^1$S and $^3$S channels of the
two-body YN interaction.

The remaining significant and challenging theoretical subject is the
consideration of the effect of the $\Lambda np$ three-body forces (3BFs)
on the $\Lambda d$ correlation function. An efficient method
for evaluating the matrix elements of the chiral $\Lambda NN$
3BFs was developed in Ref. \cite{KKM22} and benchmarked in Ref. \cite{LE25}.\\

\bigskip

{\it Acknowledgments.}
We are grateful to K. Miyagawa for his valuable discussions and comments on this study.
This work is supported by JSPS KAKENHI Grants No. JP19K03849, No. JP22K03597,
No. JP24K07019, and JP25K07301.

\appendix*
\section{Scattering wave functions in both configuration space and momentum space}
The $s$-wave radial wave functions for the $NN$ scattering are
derived from the $t$-matrices in momentum space. The single (S=0) and triplet (S=1)
wave functions with the initial momentum $p$ are
\begin{align}
 \psi_{\ell=0}^S (r;p) =& \sqrt{\frac{2}{\pi}}\left\{ j_0(pr) +\int_0^\infty p_3^2 dp_3 \sum_{\ell'=0,2}
 (\delta_{S0}\delta_{\ell'0}+\delta_{S1}) \right. \notag \\
 & \left. \times \frac{j_{\ell}(p_3 r)t_{\ell,\ell'}^S(p_3,p)}
{\frac{\hbar^2}{2\mu}p^2-\frac{\hbar^2}{2\mu}{p_3^2}+i\epsilon}\right\},
\end{align} 
where $\mu=\mu_{NN}=m_N/2$ is a reduced mass, $j_\ell$ stands for a spherical Bessel function,
and the factor $\sqrt{\frac{2}{\pi}}$ corresponds to the natural normalization of the plane wave.
The numerical integration of $p_3$ with the pole at $p_3=p$ is carried out by the subtraction method.
The wave function for the $\ell'=0$ part is expressed as follows,
employing the relation $\frac{1}{p^2-p_3^2+i\epsilon}=P\frac{1}{p^2-p_3^2}-i\frac{\pi}{2p}\delta(p_3-p)$:
\begin{align}
 & \psi_{\ell=0,(\ell'=0)}^S (r;p) = \sqrt{\frac{2}{\pi}}
 \left\{ (1-i\frac{\pi \mu p}{\hbar^2} t_{0,0}^S(p,p))j_0(pr) \right. \notag\\
  & \hspace{3em}+\left. \frac{2\mu}{\hbar^2} P\int_0^\infty p_3^2 dp_3
 \frac{ j_{0}(p_3r)t_{0,0}^S(p_3,p)}{p^2-{p_3^2}}
 \right\}.
\label{eq:a1}
\end{align}
The standard prescription for the numerical treatment of the principal-value integration is
\begin{align}
 & P\int_0^\infty p_3^2 dp_3 \frac{ j_{0}(p_3 r)t_{0,0}^S(p_3,p)}{p^2-{p_3^2}} \notag\\
 =& \int_0^\infty p_3^2 dp_3 \frac{ j_{0}(p_3r)(t_{0,0}^S(p_3,p)-t_{0,0}^S(p,p))}{p^2-{p_3^2}} \notag\\
 & + t_{0,0}^S(p,p)P\int_0^\infty p_3^2 dp_3 \frac{ j_{0}(p_3 r)}{p^2-{p_3^2}} \notag\\
 =& \int_0^\infty p_3^2 dp_3 \frac{ j_{0}(p_3 r)(t_{0,0}^S(p_3,p)-t_{0,0}^S(p,p))}{p^2-{p_3^2}} \notag\\
 & -\frac{\pi}{2}\frac{1}{r} \cos(pr)t_{0,0}^S(p,p).
\end{align}
Because the phase shift $\delta_{\ell=0}^S$ is defined as
\begin{equation}
 e^{2i\delta_{\ell=0}^S} =1-2\pi i \frac{\mu}{\hbar^2}p t_{0,0}^S(p,p),
\end{equation}
Eq. (\ref{eq:a1}) can be reduced to
\begin{align}
 & \psi_{\ell=0}^S (r;p) = \sqrt{\frac{2}{\pi}}\left\{ \frac{1}{pr}e^{i\delta_{\ell=0}^S}
 \sin(pr+\delta_{\ell=0}^S) \right. \notag \\
 & \left. +\frac{2\mu}{\hbar^2} \int_0^\infty p_3^2 dp_3
 \frac{ j_{0}(p_3 r)(t_{0,0}^S(p_3,p)-t_{0,0}^S(p,p))}{p^2-{p_3^2}} \right\}.
\end{align} 

The corresponding wave function in momentum space is expressed as follows:
\begin{align}
  \tilde{\psi}_{\ell=0}^S (p_3;p) =& \frac{\delta(p_3-p)}{p_3p}e^{i\delta_{\ell=0}^S}
 \cos(\delta_{\ell=0}^S)+\frac{2}{\pi p} e^{i\delta_{\ell=0}^S}P\frac{1}{p_3^2-p^2} \notag \\
 & +\frac{2\mu}{\hbar^2} \frac{(t_{0,0}(p_3,p)-t_{0,0}(p,p))}{p^2-{p_3^2}},
\end{align}
This is a generalized function. The $\delta$ function and the principal values are taken care of
in the integration in Eq. (\ref{eq:g3}). It is noted that $\tilde{\psi}_{\ell=0}^S (p)$ is reduced
to the free solution of $\frac{1}{p_3 p}\delta(p_3-p)$ in the case of $V_{np}=0$.

\end{document}